\documentclass[pra,aps,twocolumn,eqsecnum]{revtex4}

\usepackage{graphicx}

\begin{document}

\title{An experiment on multiple pathway quantum interference for the advanced undergraduate physics laboratory}

\author{Clark Vandam, Aaron Hankin, and A. Sieradzan}
\affiliation{Department of Physics, Central Michigan University, Mt. Pleasant, MI 48858}%

\author{M.D. Havey}%
\affiliation{Department of Physics, Old Dominion University, Norfolk, VA 23529}%
\email{mhavey@odu.edu}

\date{\today }

\begin{abstract}
We present results on a multiple-optical-path quantum interference
project suitable for the advanced undergraduate laboratory. The
experiments combine a conceptually rich set of atomic physics
experiments which may be economically developed at a technical
level accessible to undergraduate physics or engineering majors.
In the experiments, diode-laser driven two-quantum, two-color
excitation of cesium atoms in a vapor cell is investigated and
relative strengths of the individual hyperfine components in the
$6s^{2}S_{1/2} \rightarrow 7s^{2}S_{1/2}$ transition are
determined. Measurement and analysis of the spectral variation of
the two quantum excitation rate clearly shows strong variations
due to interfering amplitudes in the overall transition amplitude.
Projects such as the one reported here allow small teams of
undergraduate students with combined interests in experimental and
theoretical physics to construct instrumentation, perform
sophisticated experiments, and do realistic modelling of the
results.
\end{abstract}

\maketitle%

\section{Introduction}
Nonlinear optics has developed into a mature scientific field
\cite{he,boyd,shen}, and yet the subject continues to drive
investigations into new areas ranging from development of quantum
memories and quantum repeaters \cite{nielsen,ficek}, to novel
laser types \cite{cao} including random lasers in ultracold atom
samples \cite{kaiser}. At the same time, an important emerging
approach to physics education has been to place strong emphasis on
significant scientific research experiences for undergraduate
students.  In fact, it is not unusual for undergraduate physics
majors to start working in a research laboratory as soon as they
begin their education in the university.  In this regard, research
experiences involving laser based optical sciences are well known
to be particularly attractive to undergraduate science students.
This is in part because the existing technology is mature, is
visible, and at the same time is quite technically and
scientifically accessible to physics undergraduates.  This fosters
relatively rapid independent work in the laboratory, a definite
asset to developing interest and practical knowledge in students.

One area of nonlinear optics that is particularly accessible and
attractive as a component of undergraduate research is the
application and study of two photon processes, which include
two-photon resonant and nonresonant absorption, stimulated Raman
scattering, and a wide range of coherent processes associated with
coherent population trapping. With the pioneering experiments on
two-photon absorption within the hyperfine structure of the 5s
$^{2}S_{1/2}$ level in the sodium atom
\cite{metcalf,Biraben1974,Levenson1974}, two-photon spectroscopy
established itself as an important experimental spectroscopic
technique, routinely providing sub-Doppler resolution in atomic
level structure measurements. With the advent of relatively
economical tunable and single mode diode lasers
\cite{wieman1,wieman2}, experiments of this kind, often performed
with a single light source, have found their way into
undergraduate student laboratories \cite{Olson2006}.

By using two independently tuned diode lasers instead of one, thus
performing two-color two-quantum spectroscopy, one can add other
degrees of freedom to the experiments, these being the
polarization states of the two exciting light sources and an
adjustable energy splitting between the two quanta needed to drive
the transition. This technique has been utilized in many atomic
and molecular physics studies
\cite{Nez1993,Grove1995,Ryan1993,Bayram2006}, including the
precise measurement of relative atomic transition probabilities.
It has also been been the focus of several research projects
completed in our laboratory
\cite{beger,meyer,havey,bayram,markhotok,Bayram2006}. The
experiments were motivated primarily by the need for atomic
physics benchmarks with which to measure the precision of high
level atomic physics structure calculations, particularly in
atomic Cs \cite{Derevianko,Flambaum}.  These relativistic atomic
physics calculations, in turn, are essential to interpretation of
parity violation measurements in atomic Cs
\cite{Derevianko,Flambaum,Wood,Bouchiat,Guena,Lintz}. In the
series of measurements, two-quantum atomic transitions and their
relative strengths as a function of intermediate (virtual) level
detuning from the real atomic level were studied, and high
precision data on transition matrix elements involving S, P, and D
doublets in alkali atoms were obtained. In these experiments it
was interplay between the polarization of the exciting light
fields and the detuning from one photon resonance that led to the
high precision obtained. Physics undergraduate majors and Master
of Science students played important roles in several of these
projects.

\begin{figure}[tp]
\includegraphics{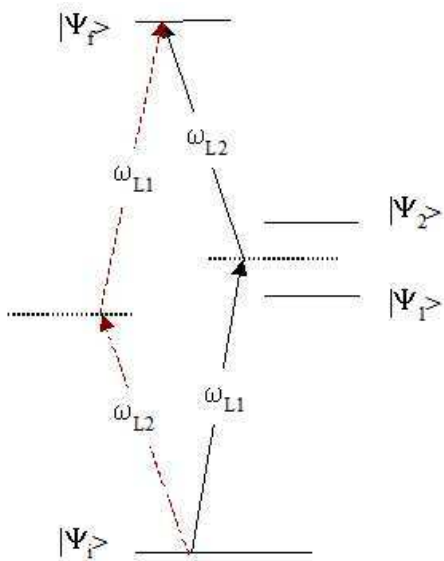}%
\caption{Generic energy level diagram of selected energetically
low-lying atomic energy levels and two-quantum excitation pathways
from an initial level to a final level. The wavefunction for the
initial level is denoted $|\Psi_i>$, while that for the final
level as $|\Psi_f>$.  The two intermediate levels are denoted
$|\Psi_m>$ with m = 1,2.  The frequencies of the two laser light
sources exciting the transition are labelled as $\omega_Lk$ with k
= 1,2. Horizontal dotted lines represent the two so-called virtual
levels for the transition. Drawing not to scale.}
\end{figure}

To illustrate the basic ideas, we turn first to Fig. 1, which
shows a generic arrangement of four energy levels, and a
two-photon excitation scheme.  In the scheme, excitation is
resonant with the two-photon transition $\emph{i} \rightarrow
\emph{f}$, but not separately resonant with either the first or
second steps of the process. Qualitatively, the two photon
process, in this case leading to population transfer from the
initial to the final level may proceed coherently through either
of the two intermediate levels. This by itself leads to
interferometric variation of the transition probability as a
function of the frequencies of the two light sources.   In
addition, the two time orders of interaction with the two light
sources are coherent with each other, and with the amplitudes for
transition through the two intermediate levels. The relative size
of these contributions depends on the detuning $\delta$ from
resonant step-wise absorption, where the two laser sources are
frequency tuned to exact resonance with the atomic level
separations. In Fig. 1, $\delta$ = $\omega_{L1} - \omega_2$,
subject to the constraint that the two photon resonance condition
is satisfied: $\omega_{L1} + \omega_{L2}$ = $\omega_f - \omega_i$.

More quantitatively, the relative polarization-dependent,
two-photon transition amplitude $M_{tp}$ is given, in the electric
dipole approximation,  by \cite{loudon}

\begin{eqnarray}
M_{tp} = \lefteqn{ \sum_{n}\left[
\frac{<\Psi_{f}|\epsilon_{1}\cdot
r|\Psi_{n}><\Psi_{n}|\epsilon_{2}\cdot
r|\Psi_{i}>}{\omega_{n}-\omega_{L1}}+
\right. }\nonumber \\
 && +  \left.
\frac{<\Psi_{f}|\epsilon_{2}\cdot
r|\Psi_{n}><\Psi_{n}|\epsilon_{1}\cdot
r|\Psi_{i}>}{\omega_{n}-\omega_{L2}}\right].
\label{Amplitude}
\end{eqnarray}

In this expression, $\epsilon_{k}$ and $\omega_{Lk}$ represent the
polarization vector and the frequency of the light source
(\emph{k} = 1,2). The index n labels the contributing intermediate
levels (in Fig. 1, n = 1,2) and their frequencies $\omega_{n}$.
For the two intermediate levels of Fig. 1, there are then four
terms in the expression for $M_{tp}$.  The $\emph{i} \rightarrow
\emph{f}$ transition probability is proportional to $|M_{tp}|^2$,
leading in this illustration to 16 contributing interfering terms.
We should point out that there are generally an infinite number of
contributions associated with off-resonance transitions to more
energetic electronic states.  These contributions are often small,
and where necessary can be estimated through known atomic
spectroscopy data.

In many applications, ground state alkali atoms are excited to one
of the higher S or D states via a two-quantum process, with an
intermediate virtual level in the vicinity of the resonance
$^{2}P_{j}$ doublet level (j = 1/2, 3/2) \cite{radzig}. Quantum
interference between transition amplitudes associated with the two
allowed excitation paths is then evident in dramatic changes of
overall transition probability, as the detuning varies. From a
fluorescence or ionization signal dependence on detuning, the
relative transition probabilities for fine structure doublets can
be determined.  The dependence on the relative polarizations of
the two light sources can lead to much greater precision in the
relative transition amplitudes, but requires very good
experimental control of the laser frequencies and polarizations.
As mentioned earlier, this approach has been used to obtain
precise relative transition amplitudes in a number of cases
\cite{beger,meyer,havey,bayram,markhotok,Bayram2006}.

We should also point out that interferences between contributing
pathways, as appear in Eq. 1, have important applications in the
broad field of quantum control of atomic and molecular processes
\cite{Brumer,Elliott,Yamazake}. In the present case, the quantum
or classical nature of the exciting fields can strongly influence
the overall transition rate \cite{loudon}, but this dependence
does not appear in the interfering terms in the configuration we
consider here. In elaborations of the scheme of Fig. 1, they can
be made to appear in interesting ways.  For example, addition of a
third field connecting one of the levels with n = 1, 2 to a yet
higher energy level can make the transition amplitude through the
n = 1,2 levels more or less distinguishable, and modify the phase
and amplitude of one or more of the interfering terms.

In this paper, we present results on a multiple path quantum
interference experiment suitable for the advanced undergraduate
laboratory. In the experiments, two-quantum, two-color excitation
of cesium atoms in a vapor cell is investigated.  In particular,
the $6s^{2}S_{1/2} \rightarrow 7s^{2}S_{1/2}$ transition is
studied, and the relative probability of transition to the final
$7s^{2}S_{1/2}$ level is measured as a function of exact frequency
offset from resonant two photon excitation.  Measurements are made
in the vicinity of the $6s^{2}S_{1/2} \rightarrow 6p^{2}P_{3/2}$
(D2) transition, in which case the interfering levels (viz. Eq. 1)
are the hyperfine components of this transition.  The observed
strong interference effects are analyzed by accounting for the
various contributing transition amplitudes and the significant
Doppler broadening of the transitions. Because of the combination
of experimental and analysis components in projects of the type
reported here, small teams of undergraduate students with combined
interests in experimental and theoretical physics can collaborate
in construction of instrumentation, performance of sophisticated
experiments, and realistic modelling of the results.

The remainder of this paper is organized as follows.  We first
provide a sketch of the theoretical results necessary for the
interpretation of the measurements and for fitting the observed
spectral variations of the two-photon transition rate data. This
is followed by a detailed description of the experimental
arrangement and protocols, and a discussion of the experimental
results.  We conclude with a brief summary and perspective on the
type of project described here.

\section{Theory}
In this section, we present some details of application of the
general expression for $M_{tp}$ to the $6s^{2}S_{1/2} \rightarrow
7s^{2}S_{1/2}$ transition in atomic Cs. This treatment consists of
two broad parts.  In the first subsection, we present an
illustrative treatment of this transition neglecting hyperfine
structure in any of the electronic levels. This allows tracing of
the essential steps in the calculation, these being the same for
the algebraically more complex case where the hyperfine splitting
is included.  This approach also turns out to be useful in its own
right, as it gives the correct result in the case where the ground
level is initially unpolarized and where the offset of the
exciting light sources is large compared to the intermediate
levels ($6p ^2P_{1/2}$ and $6p ^2P_{3/2}$ hyperfine structure. In
a second subsection, we provide an overview of a more general
approach to calculation of $M_{tp}$.  This is followed by
presentation of theoretical expressions for the
hyperfine-structure dependent spectral variations of the relative
two photon transition signals.

As is well known, atomic perturbation
theory predicts a nonzero probability for two-quantum
electric-dipole transitions between atomic states of the same
parity \cite{loudon,cohen}. Starting with an alkali atom in its
ground state, for example, higher lying S and D states may be
populated by two light sources whose frequencies add to the
frequency separation of the initial and final levels. For a
naturally broadened transition, theoretical expressions for the
transition amplitudes are readily available \cite{loudon}. These
expressions depend on the single photon transition matrix elements
between the initial and intermediate levels, and between the
intermediate levels and the final selected level.  The matrix
elements depend sensitively on the polarization of the two light
sources used. The amplitudes depend importantly also on the
detuning of the lasers from exact one plus one two photon
resonance.  With reference to Figure 2, this detuning is
represented by the offset of Laser 1 from resonance (indicated by
the position of a virtual level) through the 6p $^{2}P_{3/2}$
level. In fact, and in general, for each allowed two photon
amplitude, there are two such detunings reflecting the fact that
the two temporal orders of photon absorption are
indistinguishable. In the general case, there are also many
intermediate levels which sensibly contribute to the two photon
transition probability. However, the simplest situation occurs
when one or several energetically close terms in the summation of
the contributions dominate, while the rest contribute at the level
of often-negligible corrections. In more typical circumstances,
when several atomic levels with comparable detuning are present,
the corresponding terms in transition amplitude must be kept and
the overall sum squared to obtain the overall transition
probability.

\begin{figure}[tp]
\includegraphics{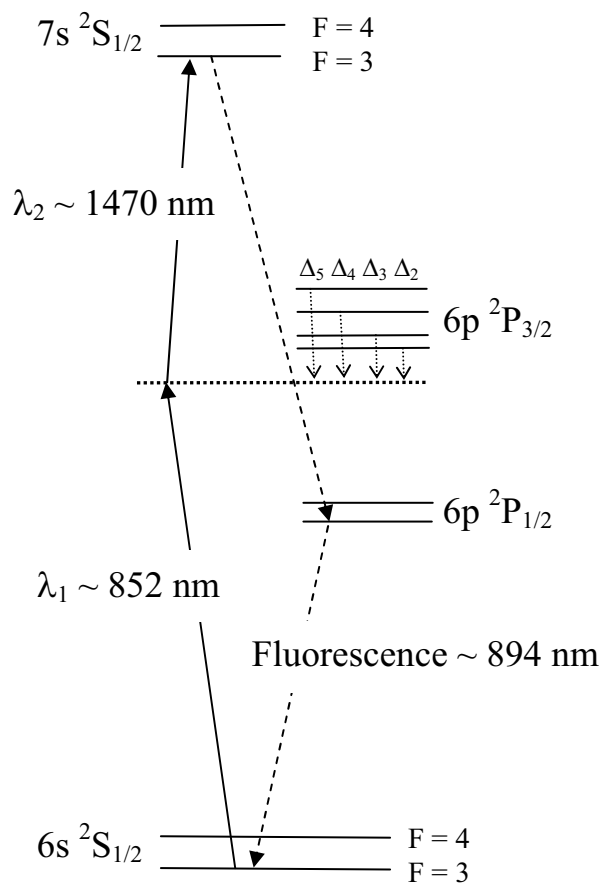}%
\caption{Schematic energy level diagram of selected energetically
low-lying atomic Cs energy levels and two-quantum excitation
pathways to the final 7s $^{2}S_{1/2}$ level.  The horizontal
dotted line represents the main contributing virtual level for the
transition. Horizontal lines schematically indicate the hyperfine
levels within each indicated atomic electronic level, while
vertical dotted lines represent the detuning of the near-resonance
virtual level from the various 6p $^{2}P_{3/2}$ hyperfine levels.
Not to scale.}
\end{figure}

To evaluate the intensity ratio for hyperfine line components in
two-quantum S-S transition, one needs to know the relevant single
photon transition matrix elements and the detuning values. It is a
useful and accessible exercise in introductory quantum mechanics,
and in angular momentum coupling rules in particular \cite{rose},
to obtain the transition amplitudes starting with an alkali atom
without spin, then include the effects of the fine structure, and
finally consider a real atom with its nuclear spin and the
associated hyperfine structure. We have performed such
calculations for all permitted S-P transitions in $^{133}Cs$,
which has a nuclear spin I = 7/2. We refer the reader to the
excellent summary of transition elements by Steck \cite{Steck}.
The various contributing hyperfine levels, for the transitions
specifically considered in the present paper, are shown
schematically in Figure 2.

In evaluating two-quantum transition probabilities between
specific hyperfine levels (e.g. from the F = 4 hyperfine component
of the $6s ^{2}S_{1/2}$ ground level to the F' = 3 hyperfine level
associated with the $7s ^{2}S_{1/2}$ electronic state), we assume
equal populations for Zeeman sublevels of the ground state and
identical reduced transition matrix elements across the hyperfine
manifold. Indistinguishable two-quantum excitation paths have
their transition amplitudes added, while transition probabilities
are summed for distinguishable paths, and the final result for the
total $F \rightarrow F'$ transition probability is obtained.  In
the following expressions, we neglect the generally interfering
contribution from the $6p ^{2}P_{1/2}$ level. We should point out
that, for lighter alkali atoms with smaller D-line fine-structure
splitting, this is not a very good approximation.  In the present
case, these variations have a negligible effect on the
experimental results. Then excluding common constant factors, the
transition probabilities, for perpendicularly polarized excitation
beams, are proportional to

\begin{eqnarray}
\nonumber P_{33} = \frac{34}{63\Delta_{2}^{2}} +
\frac{35}{128\Delta_{3}^{2}} +
\frac{125}{896\Delta_{4}^{2}} - \\
\frac{1}{6\Delta_{2}\Delta_{3}} - \frac{5}{14\Delta_{2}\Delta_{4}}
- \frac{15}{64\Delta_{3}\Delta_{4}}
\end{eqnarray}

\begin{eqnarray}
P_{43} = P_{34} = \frac{21}{128\Delta_{3}^{2}} +
\frac{119}{384\Delta_{4}^{2}} + \frac{7}{64\Delta_{3}\Delta_{4}}
\end{eqnarray}

\begin{eqnarray}
\nonumber P_{44} = \frac{161}{3456\Delta_{3}^{2}} +
\frac{2009}{9600\Delta_{4}^{2}} +
\frac{682}{675\Delta_{5}^{2}} - \\
\frac{49}{960\Delta_{3}\Delta_{4}} -
\frac{77}{270\Delta_{3}\Delta_{5}} -
\frac{77}{150\Delta_{4}\Delta_{5}}
\end{eqnarray}

where $\Delta_{F}$ represents the virtual level detuning from
individual hyperfine structure sublevels of the $6p^{2}P_{3/2}$
cesium level. These detunings are indicated schematically in
Figure 2. Note that, in practical terms, the detuning of Laser 1 (see Figure 2),
labelled for instance as x, is referenced to the frequency of the
hyperfine-average $6s ^{2}S_{1/2} \rightarrow 6p ^{2}P_{3/2}$
resonance transitions. Individual $\Delta_{F}$ are then expressed
in terms of x using the known hyperfine level locations
For example, $\Delta_{5}$ = 3758 MHz  + x, for excitation from F =
4.

In the limit where all detuning become large compared to
the hyperfine splittings, and all $\Delta_{F}$ are nearly
identical, the ratios of transition probabilities become constant,
and are given by $P_{43}/P_{44} = 7/5$ and $P_{34}/P_{33} = 3/1$.
In general the spectral dependence of the ratios depends in a
complicated way on detuning, but are in any case given by Eq. 2.1
- 2.3, so long as the detunings are all larger than the Doppler
and natural widths of the individual hyperfine transitions. These
results are valid for $^{133}Cs$ only, but similar calculations of
relative transition probabilities can be easily extended to other
systems. Strictly speaking, the above expressions are correct only
when the individual transition amplitudes scale as $1/\Delta$, and
do not directly apply when the detuning becomes on the order of
the natural width of any of the individual hyperfine resonances.
How wide a spectral range needs to be so that these expressions
apply depends on the natural width of the transitions and on the
temperature of the sample. For experiments performed with
cold atoms \cite{metcalf}, the natural width of a  few MHz would
determine the "resonance region". In the case considered here,
where a cesium vapor was kept at 323 K, the Doppler broadened
resonance frequencies spread over hundreds of megahertz and to
produce theoretical curves to compare with experimental intensity
ratios, one needs first to integrate the transition probability
expressions over the Doppler distributed resonance frequencies,
and then make proper comparisons between theory and experiment
outside the Doppler width range of about 500 MHz around the center
of each resonance. Results obtained from such a procedure are used
for comparison with experimental data in the following sections of
the paper.

\section{Experimental Description and Protocols}
The experimental arrangement is shown schematically in Figure 3.
Excitation of Cs to the $7s ^{2}S_{1/2}$ level is achieved by
means of two independently tuned, extended-cavity diode lasers.
Each laser operates on a single longitudinal mode, and provides a
few milliwatts of output power. A fraction of this intensity, depending on detuning from hyperfine
resonance, was used to excite the Cs atoms to the $7s ^{2}S_{1/2}$ state.  The linearly laser beams pass
through optical isolators, after which they are polarized
orthogonal to each other.  They are then directed, in a
counter-propagating fashion, through a Pyrex glass cell with flat
windows containing saturated cesium vapor at an equilibrium cell
temperature stabilized to 323(1) K, maintaining a cesium vapor
pressure of about $10^{-5}$ Torr. To avoid two-photon transition
saturation, and preserve good collimation of the beams, no
focusing lens is used in either of the beams. In addition to this
precaution, neutral density filters are inserted in the beams when
either laser frequency approaches single photon resonance.

\begin{figure}[tp]
\includegraphics{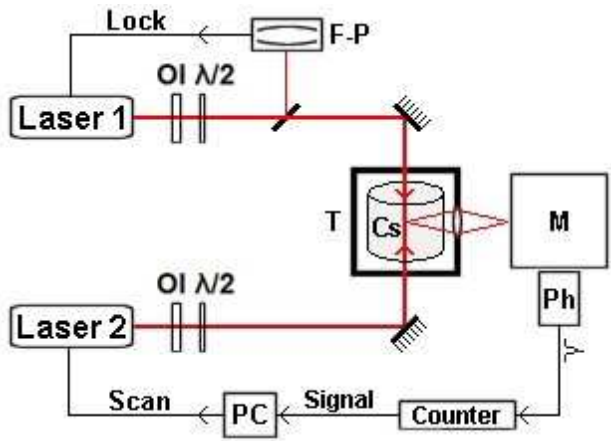}%
\caption{Schematic diagram of the experimental apparatus.
Abbreviations in the figure correspond to OI (optical isolator),
$\lambda/2$ (half-wave retardation plate), T (sample cell oven), M
(monochromator), Ph (photomultiplier tube), PC (data acquisition
computer, F-P (Fabry-Perot interferometer), and Cs (the atomic Cs
vapor cell). }
\end{figure}

Atomic population in the $7s ^{2}S_{1/2}$ level is monitored by
the cascade fluorescence through the $6p ^{2}P_{1/2}$ level at 894
nm (the Cs D1 line). The fluorescence light from the cell is
spectrally filtered by a 0.5m double monochromator, and detected
with a gallium arsenide photocathode photomultiplier operating in
a photon-counting mode.

The first laser has its wavelength set at the beginning of each
run. The virtual level detuning from the $6p ^{2}P_{3/2}$ level
resonance is determined by counting transmission cycles in a
confocal Fabry-Perot interferometer (which has a free spectral range of 2 GHz), observed during the laser
frequency adjustment starting at the point of maximum absorption
of an attenuated beam in an auxiliary room-temperature Cs
reference cell. Comparison with a more elaborate saturated
absorption scheme allowed us to determine that such a simple
reference results in laser frequency determination and
reproduction with an accuracy of about $\pm$ 60 MHz, more than
adequate for the present experiment.

After the initial adjustment, the frequency of laser 1 is
electronically locked to a transmission peak in the
interferometer. The frequency stability of the lock, with the
interferometer contained in a thermally insulated enclosure
without active temperature stabilization, is 50 MHz or better for
the typical one-hour duration of a run. The overall quadrature
laser 1 frequency determination error is then $\pm$ 80 MHz. The
second diode laser, laser 2, has its frequency scanned in a narrow
range over which two-quantum resonances are observed. Relative
intensities of the four excitation/fluorescence peaks, which
correspond to different F, F' combinations in the two-quantum
$6s^{2}S_{1/2} \rightarrow 6p ^{2}P_{3/2} \rightarrow
7s^{2}S_{1/2}$ transition, are measured after a small background
signal subtraction. As the four lines are of nearly identical
width, either the integrated area under the line, or the peak
amplitude, can in principle be utilized as a measure of line
intensity. Although both quantities produce similar results for
the various intensity ratios, we choose the amplitude comparison
to avoid distortion of the ratios associated with residual
non-uniformity in the laser frequency scan rate.

Uncertainties in the peak ratio determinations comes from a
combination of photon counting statistics, fluctuations in the
thermal lab environment, and laser power variations. While beam
power just before the cell show very small changes ($<0.1$
percent), the cell entrance window, which has no antireflection
coating, can act like an etalon with reflection/ transmission
varying somewhat with laser frequency. A simple estimate, as well
as auxiliary experimental tests, show that with a characteristic
laser 2 frequency change of 2 GHz between peaks corresponding to
different F', the cell window transmission never changes by more
than about 1 percent.  The total uncertainties in the ratios are
obtained by adding contributions in quadrature.

\begin{figure}[tp]
\includegraphics{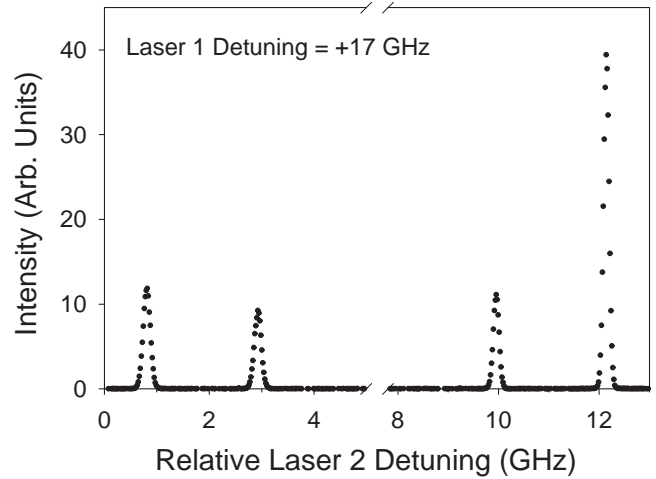}%
\caption{Illustrative scan of the four hyperfine resonances
associated with far-off-intermediate-level resonance of the
$6s^{2}S_{1/2} \rightarrow 7s^{2}S_{1/2}$ transition. The relative
detuning label on the abscissa indicates that the absolute zero of
the scan is not specified, but that the relative resonance
separations correspond to the detuning differences given in the
figure. On the other hand, the detuning indicated in the graph
body refers to the hyperfine-averaged $6s^{2}S_{1/2} \rightarrow
6p ^{2}P_{3/2}$ transition frequency (see text).}
\end{figure}

\begin{figure}[tp]
\includegraphics{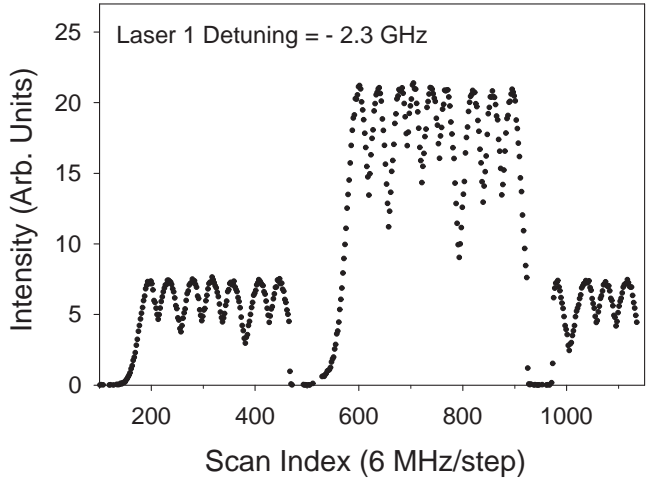}%
\caption{Typical data showing repeated scans of two hyperfine
levels, in this case the transitions initiating in the lower F = 3
level. This data illustrates the temporal stability of the
apparatus for the purposes of measuring relative intensity ratios.
The scan index refers to a single step in the frequency scan
(approximately 6 MHz) as the laser is scanned back and forth over
the hyperfine resonances. The detuning indicated in the graph body
refers to the hyperfine-averaged $6s^{2}S_{1/2} \rightarrow 6p
^{2}P_{3/2}$ transition frequency (see text).}
\end{figure}

\section{Results and Discussion}
Initial tests showed two-quantum-induced fluorescence signals
growing strongly with decreasing detuning, in agreement with
theoretical predictions. Since the absolute fluorescence signal
intensity is not very reproducible from day to day, we keep
fluorescence signals roughly constant (for example by adjustment
of the monochromator slit width), and concentrate on the relative
intensity of the various hyperfine transition resonances. With the
laser 1 frequency set and laser 2 scanned, four fluorescence peaks
are observed corresponding to different F = 3,4 to F' = 3,4
combinations in the $6s^{2}S_{1/2} \rightarrow 6p ^{2}P_{3/2}
\rightarrow 7s^{2}S_{1/2}$ two-quantum transition.  A
characteristic scan over these resonances are illustrated in
Figure 4. We should emphasize here that, in all experimental
results, detunings refer to the hyperfine-averaged $6s^{2}S_{1/2}
\rightarrow 6p ^{2}P_{3/2}$ transition frequency \cite{Steck}.
Since the hyperfine energy separation in the lower ground level is
several times larger than in the final $7s ^{2}S_{1/2}$ level -
the spectral signal consists of two doublets separated by the
ground level hyperfine splitting. The intensity ratios are
carefully determined for pairs of transitions originating from a
common lower level, i.e. those starting at F = 4, and then those
starting at F = 3. Such ratios are insensitive to the population
in the ground state hyperfine structure sublevels. As already
mentioned, the amplitudes of the fluorescence peaks rather than
their integrated areas are used as transition probability measures
and their ratio taken. Several consecutive scans over each line
center are made, in order to assure consistency of the
measurements. Occasional records, showing noticeable drifts in
values within the series are rejected. Typical scan results are
illustrated in Figure 5. The peak regions are fitted to a
parabolic line shape and the maximum signal values determined by
the fit. After subtraction of the background, which consists of
dark counts and stray light, the resulting derived values are used
for amplitude ratio evaluation. For larger detunings (hyperfine
structure splitting $<<$ detuning $<<$ fine structure splitting)
these values are expected to approach 1.4 and 3, as mentioned
earlier. Our experimental data illustrating these effects is shown
in Fig. 6 and Fig. 7.

\begin{figure}[tp]
\includegraphics{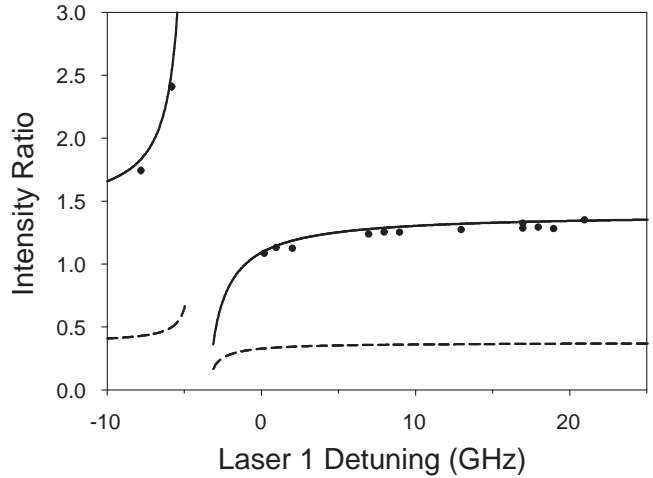}%
\caption{Spectral variations of the measured intensity ratios of
the F = 4 $\rightarrow$ F' = 3 to the F = 4 $\rightarrow$ F' = 4
hyperfine transition. The solid line represents the theoretical
prediction, calculated as described in the text.  The dashed curve
represents the theoretical prediction, but ignores interference
among the two photon hyperfine transition amplitudes.  The laser 1
detuning in this figure is referenced to the hyperfine averaged
line center.}
\end{figure}

\begin{figure}[tp]
\includegraphics{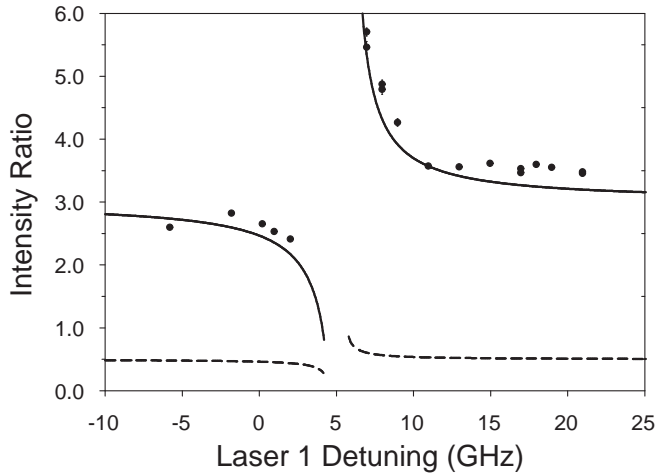}%
\caption{Spectral variations of the measured intensity ratios of
the F = 3 $\rightarrow$ F' = 4 to the F = 3 $\rightarrow$ F' = 3
hyperfine transition. The solid line represents the theoretical
prediction, calculated as described in the text.  The dashed curve
represents the theoretical prediction, but ignoring interference
among the two photon hyperfine transition amplitudes. The laser 1
detuning in this figure is referenced to the hyperfine averaged
line center.}
\end{figure}

When the laser 1 detuning from a resonance transition becomes
comparable with the $6p ^{2}P_{3/2}$ hyperfine splitting the
ratios may deviate from the asymptotic values substantially.
Transitions populating the F' = 4 level dominate over those ending
in the F' = 3 level when the virtual level is on the higher
frequency side of the $6p ^{2}P_{3/2}$ level, and vice versa,
while the transition amplitude interference causes the intensity
ratios to strongly tend in opposite directions when resonance is
approached. In general, although a more complicated behavior of
the probability ratios are expected in, for example, an ultracold
(due to resolution of individual intermediate hyperfine levels)
atomic gas, quantum interference effects in the transition
probabilities are clearly demonstrated here as well. This is
emphasized by the dashed lines in the figures. These lines
correspond to ignoring interferences between the hyperfine
transition amplitudes, and directly combining transition
probabilities.

Although the experimental data follow quite closely the trends
predicted on the basis of the simple model presented here, there
are systematic deviations of a few percent between the
experimental data and the theoretical predictions. Several
possible sources of discrepancy were considered.  First,
considering the theoretical model used, we emphasize that the main
approximations are (a) neglect of contributions of the reverse
time-ordering of the the two photon transition amplitude, and (b)
neglect of transition amplitudes via the energetically distant $6p
^{2}P_{1/2}$ level.  Estimates readily show that these contribute
negligibly, for the detunings considered here, and at the level of
the precision of the measurements. Far off resonance transitions
due to more energetically distant np multiplets similarly
contribute at a negligible level in comparison with the
discrepancies.  It further seems unlikely that there are
significant variations of the reduced transition matrix elements
on the scale of the present experiments. A number of experimental
artifacts were also considered. These include imperfectly
perpendicular laser beam polarizations and saturation effects in
the two photon transition rate. However, it proves difficult to
reconcile the discrepancies within reasonable estimates of these
effects.  The most likely explanation for the few percent
differences between the measurements and the theoretical
predictions comes from effects of the broad spectral profile of
the diode lasers used in the experiments.  Although the main
output of the diode lasers is generally within a bandwidth $\sim$
MHz, there is a weak and broad spectral plateau. This plateau can
be spectrally structured because of low, but not negligible levels
of mode competition in the external cavity diode lasers,
especially at the spectral edges of the tuning range of the
lasers.  Mode instability in fact becomes quite evident for the
nominal 1470 nm diode laser at detunings somewhat outside the
range of than those used in the present experiments.  Quantifying
these issues is challenging without recourse to a near infrared
sensitive spectrum analyzer, but the scale of the instabilities we
observe is in fact consistent with the spectral trend and size of
the observed discrepancies.

\section{Pedagogical Perspectives}
The research described in our previous reports
\cite{beger,meyer,havey,bayram,markhotok,Bayram2006}, and in the
present paper, form the basis for an excellent set of projects
suitable for advanced undergraduate physics laboratories.  These
projects are based on atomic two-photon, two-color experiments in
which quantum interferences play an important role in the observed
transition rate.  For these experiments there are two types of
interference effect.  In one, the various amplitudes for
intermediate atomic levels contribute coherently to the total
transition probability.  In the second, the two time-orderings of
photon absorption also contribute coherently.  In many cases, one
of these orderings dominates over the other.

In the work reported here, and in our previous measurements of
this type, we have concentrated on studies of alkali atoms.
However, this is not in the least a requirement for such studies,
and a variety of atomic species, and many molecular ones as well,
could be used to provide a variety of new studies to serve as
undergraduate research experiences.  As such experiments have not
been systematically extended to the molecular domain, the roles of
multiple rovibrational interferences provide a rich area for
interesting, accessible, and original research suitable for
undergraduates.  In general, the results would also be of
sufficient interest to be publishable in traditional physics
journals.  The projects combine the desirable aspects of
experimental technical accessibility, clear physical
interpretation, and theoretical analysis within the range of many
advanced physics or engineering students.  As the theoretical
results can be readily parametrized in terms of combinations of
atomic or molecular electric dipole transition moments,
opportunities also exist for data analysis and fitting.  All
together, such projects are quite suitable for a single student
working with involved faculty, or small teams of advanced
undergraduates working together by bringing different skills and
inclinations to the projects. It has been our experience with
undergraduate and graduate students involved in similar
experiments that accurate predictions of line intensity ratios and
their variations with detuning can impress even naturally
skeptical bright undergraduate physics and engineering students.
They also come to appreciate the opportunity to see quantum
mechanics going to work and producing excellent results which can
be experimentally verified in a student level experiment.

\section{Conclusions}
We have presented measurements of relative transition
probabilities for various hyperfine transitions in atomic Cs. The
measurements are interpreted through theoretical expressions valid
for the spectral detuning range outside the Doppler widths of the
atomic transitions, where very good agreement between experiment
and theory is obtained.  The experimental approach and theoretical
results are presented in sufficient detail to guide development of
possible undergraduate research projects focussed towards similar
atomic or molecular systems.

\section{Acknowledgments}
This work is supported by Central Michigan University and by the
National Science Foundation (NSF-PHY-1068159 and NSF-PHY-1606743).

\end{document}